\documentclass{article}
\usepackage{amsmath,amssymb,graphicx,mlspconf,url}
\usepackage{wrapfig}
\usepackage{multicol, blindtext}

\def\x{{\mathbf x}}
\def\y{{\mathbf y}}

\def\R{{\mathbb{R}}}

\newenvironment{packed_enumerate}
{\vspace{-0.5em}
\begin{enumerate}
\setlength{\itemindent}{-10pt}
\setlength{\itemsep}{-2pt}}
{\end{enumerate}\vspace{-0.5em}}

\title{Acoustic Scene Classification: a competition review}
\name{\begin{tabular}{c}Shayan Gharib$^{1}$, Honain Derrar$^{1}$, Daisuke Niizumi$^{2}$, Tuukka Senttula$^{1}$, Janne Tommola$^{1}$, Toni Heittola$^{1}$,\\ Tuomas Virtanen$^{1}$, Heikki Huttunen$^{1}$\end{tabular}}
\address{$^{1}$Lab. of Signal Processing, Tampere University of Technology, Tampere, Finland\\
$^{2}$Balmuda Inc., Tokyo, Japan\thanks{This work was partially funded by the Academy of Finland
project 309903 CoefNet and ERC Grant Agreement 637422 \mbox{EVERYSOUND}. Authors also thank CSC--IT Center for Science for computational resources.
}}

\begin{document}
%

\maketitle
\begin{abstract}
In this paper we study the problem of acoustic scene classification, \textit{i.e.,} categorization of audio sequences into mutually exclusive classes based on their spectral content. We describe the methods and results discovered during a competition organized in the context of a graduate machine learning course; both by the students and external participants. We identify the most suitable methods and study the impact of each by performing an ablation study of the mixture of approaches. We also compare the results with a neural network baseline, and show the improvement over that. Finally, we discuss the impact of using a competition as a part of a university course, and justify its importance in the curriculum based on student feedback.

\end{abstract}
%
%
\section{Introduction}
\label{sec:intro}
Humans are capable of understanding the environment by listening to the sounds surrounding them. 
One of the major research areas in computational auditory scene analysis (CASA) is acoustic scene classification (ASC) \cite{hanconvolutional}. ASC attempts to classify digital audio signals into mutually exclusive scene categories,  which  can be \emph{e.g.,} an indoor environment (such as \textit{home}) or an outdoor environment (such as \textit{park}) \cite{Barchiesi2015}. 

It should be noted that this problem is different from a sound event detection (SED) task. In particular, SED attempts to classify, and detect the start and end points of sound events. Moreover, sound events may be overlapping unlike the acoustic scene classes, which are mutually exclusive.
ASC can be applied in many areas; including mobile robot navigation systems \cite{Chu2006} and context-aware devices \cite{schilit1994context}, such as an automatically mode-switching smart phones according to the current acoustic environment \cite{Eronen2006}. 

In the sequel, we describe a pool of methods for solving the ASC problem, crowdsourced during a two-month competition organized as a mandatory part of a graduate level university course with almost 200 students. Moreover, the competition was organized on the Kaggle InClass platform and was open for everyone, thus attracting talented researchers also outside the university student community. Here, we describe the common approaches that seemed to be of greatest benefit to the top teams, and attempt to synthesize a complete framework of ASC system adopting the best practices of this large crowd of expert participants.


The rest of the paper is organized as follows. In Section~\ref{sec:format}, we will review related literature, including the classical Hidden Markov Model based approaches all the way until more recent deep learning based works. In Section \ref{sec:Dataset}, we will describe the competition data, specifically tailored for this purpose. Section \ref{sec:Methods} presents the collection of successful methods, ranging from data augmentation to ensemble averaging, deep learning and semi-supervised training. Finally, in Section \ref{sec:Education}, we discuss the importance of hands-on experience in education in general, and the role of a competition as part of the computer science curriculum in particular.

\section{Related work}
\label{sec:format}

Given the huge variability of acoustic environments, the practical approach is to gather a data set with a finite set of relevant categories and collect audio recordings from each. The resulting supervised classification task \cite{Barchiesi2015} is then well defined assuming the relevance and validity of the data collection. As usual, compared to the manual engineering of acoustic indicators, learning algorithms can better handle the vast complexity and variability within the data \cite{heittola2018machine}. 

One of the earliest approaches to ACS \cite{sawhney1997situational} involves the use of feature extraction techniques used earlier in speech analysis and auditory research, coupled with Recurrent Neural Networks (RNNs) and a K-Nearest-Neighbor criterion to construct a mapping between the feature space and the sound event category. Later, researchers have used Hidden Markov Models (HMM) that allow to take into account also the temporal unfolding of events within a sound sample. This is important as using the temporal context of sound spectral components clearly improves recognition accuracy. 




More recently, with the development of deep learning \cite{lecun2015deep}, neural network approaches in general and convolutional neural networks (CNNs) in particular have become the dominant research track in the field of ASC. For example, all of the top 10 submissions in competitions such as the DCASE \cite{DCASE2017challenge} are based on a CNN-based classifier either alone or in conjunction with other techniques. CNNs have enabled researchers to obtain great results in image classification tasks such as in \cite{NIPS2012_4824,simonyan2014very}. Following this success, also audio recognition has been revolutionized by their strength. 


Competitions and community campaigns are of great importance in attracting talented researchers and pushing the state-of-the-art methods further within a research field. In this regard, DCASE challenge \cite{mesaros2018detection} is one of the competitions that is active on related fields to computational analysis of sound events and scene analysis, namely acoustic scene classification, sound event detection, and audio tagging. It started in 2013 including two tasks and 18 international participants, and has grown to already 87 teams in 2016 with 4 different tasks. Another benefit of competitions such as DCASE is to provide a fair comparison, and a baseline dataset for researchers to work with, and make the results of their studies more concrete for others to compare with. 

Apart from DCASE, music information retrieval evaluation exchange (MIREX) is another campaign which has been active for more than a decade, and covered different tasks within the field of music information retrieval \cite{downie2008music}.

Even outside of research community, platforms such as Kaggle, CodaLab or CrowdAI have attracted research institutions and companies to publish state-of-the-art challenges such as the recent TensorFlow Speech Recognition challenge organized by Google. Moreover, this opportunity is a beneficial practice not only for researchers but for people from outside of research community as well to extend and share their knowledge in a publicly accessible website. 

\section{Competition Setup and Data}
\label{sec:Dataset}

The competition was organized in the context of a two month graduate machine learning course in January-February 2018 on the Kaggle InClass platform, which is free for educational use. The course had over 200 participants that were split into groups of four students each (total 45 teams). Moreover, the competition was open for everyone, and eventually 69 teams participated; one third of all teams outside the university. The dataset (described below in more detail) features 15 ASC classes, and the performance metric for ranking the teams was the prediction accuracy (\textit{i.e.,} how many percent of the audio samples were predicted correctly). In total, there were over 700 submissions from the teams.

We arranged a new dataset, TUT Acoustic Scenes 2017 Features\footnote{https://doi.org/10.5281/zenodo.1324390} for crowdsourcing new approaches to acoustic scene classification. The goal was to have an easy to use, yet descriptive, dataset that would enable large scale adoption. Namely, many of the largest datasets (\textit{e.g.,} Google AudioSet) are challenging for a large scale adoption due to their sheer size and their dynamic nature (a collection of references to audio files, which may disappear over time).

The dataset contains similar material to TUT Acoustic Scenes 2017 dataset \cite{tut-acoustic-scenes-2017-development,tut-acoustic-scenes-2017-evaluation}, used in the DCASE2017 acoustic scene classification challenge \cite{DCASE2017challenge}. Both datasets are composed from same pool of original audio recordings, but exact audio segments selected for the datasets differ. 

The acoustic material consists of binaural recordings from 15 acoustic scenes: \textit{lakeside beach, bus, cafe/restaurant, car, city center, forest path, grocery store, home, library, metro station, office, urban park, residential area, train,} and \textit{tram}. Multiple recordings of 3-5 minutes length were captured from each scene class, and each recording was done in a different location to ensure high acoustic variability. Complete details on data recording procedure can be found in \cite{Mesaros2016_EUSIPCO}, although it should be noted that we only use a subset of the features of the complete set.

In order to simplify the adoption of the dataset, we decided to exclude the pipeline for feature extraction from raw audio, and release the dataset only as acoustic features. More specifically, we extracted the log mel-energy features in 40 mel bands for each in 40~ms window (50\% hop) for each 10-second long audio segment. Each 10-second recording is thus represented as a $40\times 501$ feature matrix (40 frequency bins from 501 time points). The log mel-energy features were selected as they have shown to be among the most informative for this type of classification problem (see \textit{e.g.,} \cite{Cakir2017}).

\subsection{Development and evaluation sets}

The dataset consists of \textit{development} set and \textit{evaluation} folds. The development set was released with ground truth, whereas the evaluation set was released without true class labels. The participants may submit their predictions for the evaluation set two times each day and the competition platform evaluates the accuracy. Moreover, the evaluation set is divided into \textit{public leaderboard} and \textit{private leaderboard} subsets (in Kaggle terminology) with a 50/50 ratio. The participants see their score for the public leaderboard subset immediately; but the final team ranking is based on the hidden private leaderboard subset, whose results are disclosed only when the competition ends. The three folds---development (75\%), public leaderboard (12.5\%) and private leaderboard (12.5\%)---correspond to the more common concepts of training, validation and test folds with the intention to discourage optimization against the test set. In absolute quantities, the development set contains 4500 audio segments with a length of 10 seconds, totaling 750~minutes of audio. The evaluation set contains 1500 10-second audio segments, 250~minutes of audio in total.  

\subsection{Material selection}

Acoustic scenes have large variability in the acoustic properties, and due to limited amount of recordings and recording locations, extra care has to be taken to make the sets of the dataset balanced. The equally balanced sets are essential for making the development process deterministic for the participants; the system should perform relatively equally in development set, in public leaderboard subset, and in private leaderboard subset.

Original recordings were averaged into single channel and cut into 10-second segments. Segments originating from the same recording were assigned into same set, either development or evaluation, to make evaluation setup more realistic. Our segment selection procedure was done independently for each scene class. In this process a large number of randomly selected set candidates were created and ranked based on how acoustically similar segments were selected to the sets (development and evaluation). The candidate sets were created by first randomly assigning recordings to sets and then randomly selecting segments from these recordings to fill the target number of segments for each set. For the development set, 300 segments were selected per scene class, and for evaluation set 100 segments.  The final sets were selected randomly among the top 25\% with best acoustic similarity. 

Acoustic similarity between the sets was determined by training a Gaussian Mixture model (GMM) for each set and measuring the Kullback-Leibler divergence between them. First, mel-frequency cepstral coefficients (MFCCs) were extracted in 40~ms windows (50\% hop size) for each segment in the sets. These features were  then aggregated over one second analysis window (50\% hop size) by calculating mean and standard deviation within the analysis window. The aggregated features were pooled separately for the development and evaluation sets, and a Gaussian Mixture model (GMM) was trained for each (32 components). Distance between these two GMMs was then calculated using the empirical symmetric Kullback-Leibler divergence \cite{Virtanen2007b,Heittola2014_EURASIP}.

Instead of the within-sequence accuracy, we wanted to measure the inter-sequence generalization as well; \textit{i.e.,} how well the the approaches generalize to completely unseen sequences. To this aim, each sequence is included in exactly one of the three folds. Also the sequence index for each development sample was provided in order to enable local inter-sequence benchmarking with the training data for the participants.

\section{Methods}
\label{sec:Methods}
In this section, we describe a collection of successful methods used  by the top teams of the competition and assess their importance in the final accuracies.

\subsection{Data Augmentation}

\begin{figure}[tb]
  \centerline{\includegraphics[width=9cm]{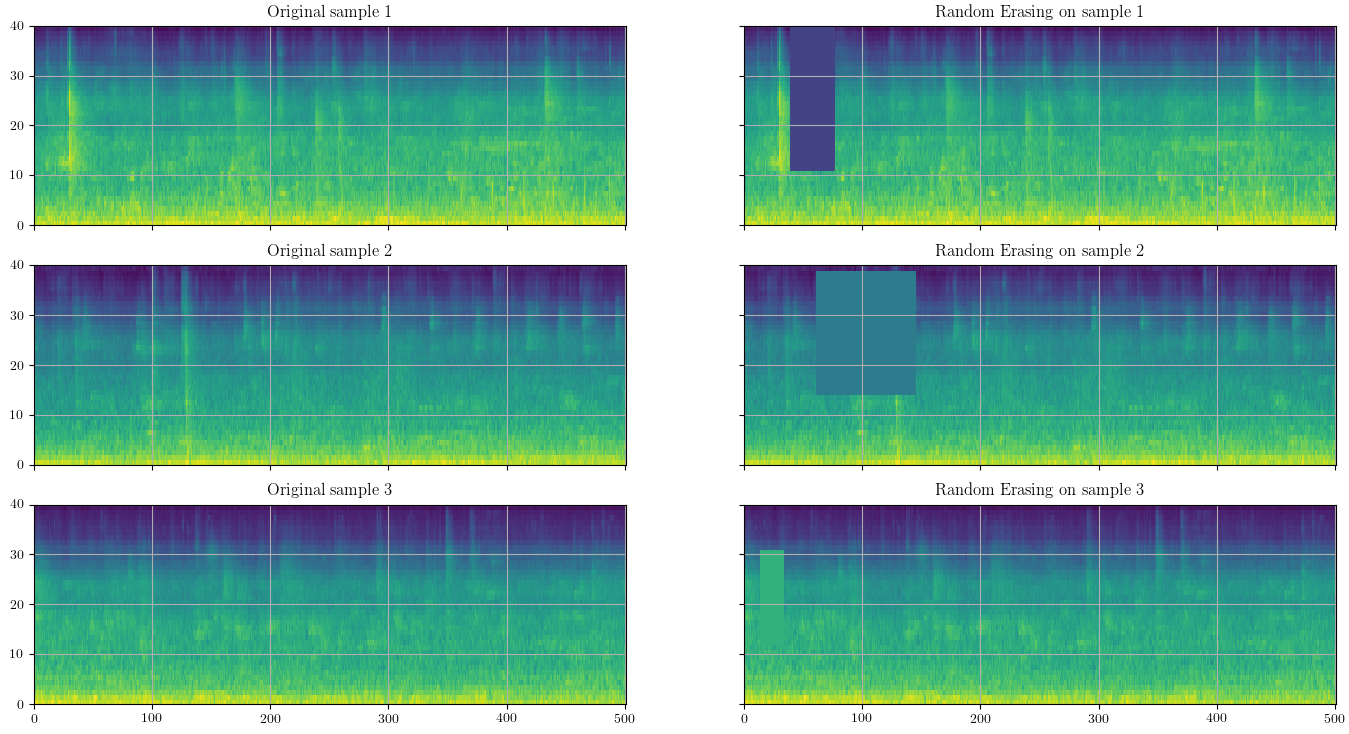}}
  \caption{Illustration of the effects of Random Erasing.}
  \label{Fig. 1}
\end{figure}

\begin{table}[tb]
\caption{Public scores obtained with random erasing and mixup augmentation with and without cyclic learning rate (CLR) scheduling.}
\vspace*{0.2cm}
\centering
\resizebox{\columnwidth}{!}{%
\begin{tabular}{l||c c c c c}
\hline\hline
\textbf{Combination} & \textbf{CLR} & \textbf{\begin{tabular}[c]{@{}l@{}} Random \\  erasing\end{tabular}} & \textbf{Mixup} & \textbf{\begin{tabular}[c]{@{}l@{}} Public \\  leaderboard\end{tabular}} &  \textbf{\begin{tabular}[c]{@{}l@{}}Delta from \\  baseline \end{tabular}} \\
\hline\hline
Baseline  & No & No & No & 75.33 \% & 0.0 \\ 
Baseline  + CLR & Yes & No & No & 72.13 \% & -3.2 \\ 
\begin{tabular}[c]{@{}l@{}} Baseline + \\   Random erasing\end{tabular}  & No & Yes & No & 75.47 \% & +0.13 \\ 
Baseline  + Mixup & No & No & Yes & 76.27 \% & +0.93 \\ 
\begin{tabular}[c]{@{}l@{}} Baseline + \\   All but CLR\end{tabular}
& No & Yes & Yes & 76.40 \% & +1.07 \\ 
Baseline  + All & Yes & Yes & Yes & \textbf{77.07 \%} & +1.73 \\ 
\hline\hline
\end{tabular}
}%
\label{table:1}
\end{table}

One of the challenges related to deep neural networks is the large amount of data that they require to generalize properly. In order to address this, one may use different regularization approaches ($\ell_1$ and $\ell_2$ penalty or dropout \cite{srivastava2014dropout}) or data augmentation. 

In this regard, the top teams of the competition used two key augmentation approaches:  \textit{random erasing} \cite{zhong2017random} (closely related to \textit{cutout} \cite{devries2017improved}), and \textit{mixup} \cite{zhang2018mixup}. As in all augmentation, the idea behind both approaches is to synthesize new (hopefully realistic) samples from the training data to force the model to better learn the natural variability within the data.

\begin{figure}[tb]
  \centerline{\includegraphics[width=9cm]{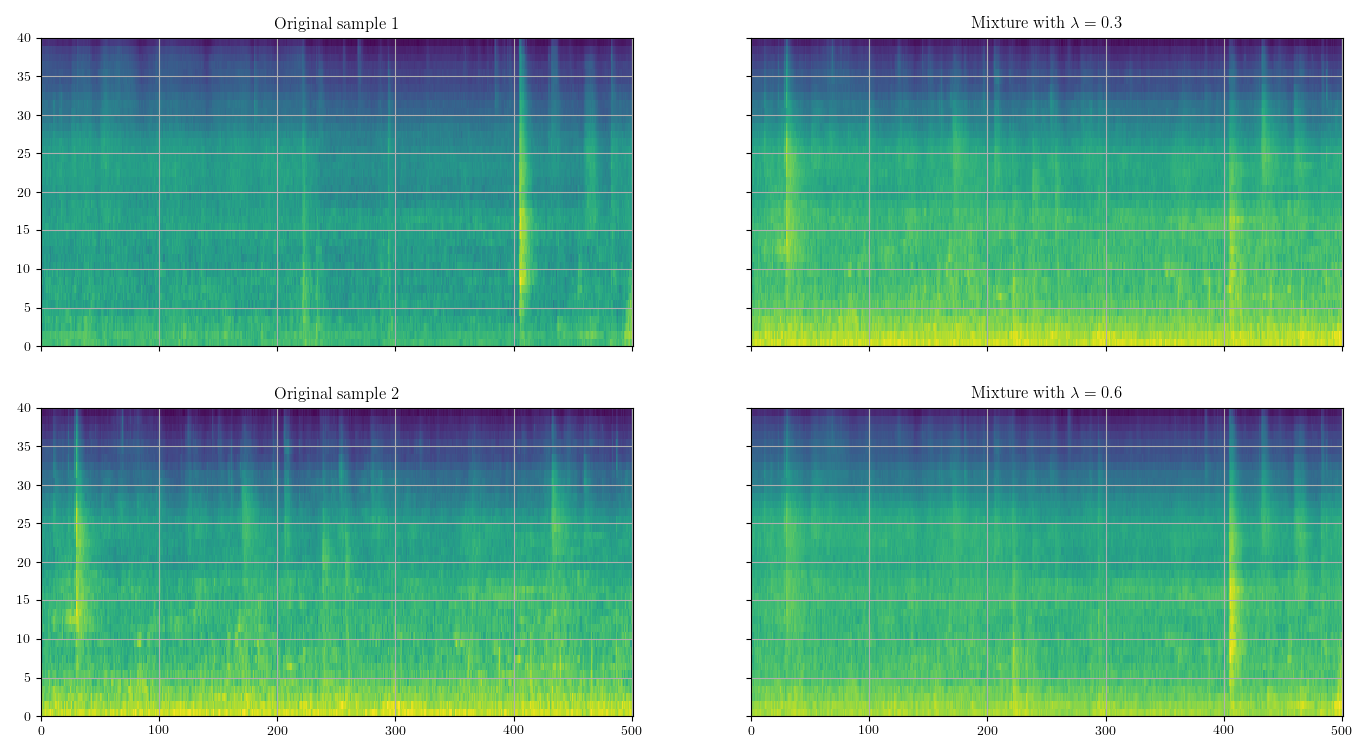}}
  \caption{Illustration of mixup augmentation. Left: Two original mel-spectrogram samples from the training set. Right: two mixtures of the original samples.}
  \label{Fig. 2}
\end{figure}

Augmentation by random erasing (and cutout) is inspired by dropout regularization: Random erasing inserts blank rectangles into the two-dimensional spectrogram (or image) at a randomly chosen size and location. The Random erasing was successfully used by the winning team, although the improvement is not as significant as that of other augmentation techniques. However, due to its simplicity it is easy to implement and make the model more robust against noise. Examples of random erasing effect are shown in Figure \ref{Fig. 1}.



A more important data augmentation technique is \textit{mixup} \cite{zhang2018mixup}. In contrast to the aforementioned approaches, mixup merges training samples together instead of just distorting individual training samples. 
More specifically, consider two randomly chosen training samples $\x_i\in\R^P$ and $\x_j\in\R^P$ with their corresponding class labels $\y_i\in \{0,1\}^C$ and $\y_j\in \{0,1\}^C$ in one-hot-encoded representation with $P$-dimensional features and $C$ classes. Then, mixup constructs a new training sample $\overset{\sim}{\x}\in\R^P$ with target $\overset{\sim}{\y}\in\R^C$ as follows:
\begin{align*}
\overset{\sim}{\x} &= \lambda \x_{i} + (1-\lambda)\x_{j},\\
\overset{\sim}{\y} &= \lambda \y_{i} + (1-\lambda)\y_{j},
\end{align*}
where $\lambda\sim Beta(\alpha,\alpha)$ with hyperparameter $\alpha > 0$. 
This way we create mixtures of \textit{both input and output} ending up at non-binary interpolated targets $\overset{\sim}{\y}$.

\par In Table \ref{table:1}, we compare the effects of the different data augmentation techniques on our baseline model which is a CNN based on the Alexnet \cite{NIPS2012_4824} architecture. 
In addition to the augmentation techniques, we also evaluate the effect of using cyclic learning rate scheduling \cite{smith2017cyclical}---a significant component used by the best teams.

As shown in the table, using random erasing and mixup without cyclic learning rate (CLR) improves the accuracy score by approximately 1\% over the base-line model. When augmentation is combined with CLR the accuracy is improved an additional 0.7\%. 
\subsection{Model Fusion}
\label{ssec:cvmv}
Many of the best teams used model fusion to combine different classifiers.
A novel idea discovered during competition is to integrate model fusion with $K$-fold cross-validation. To tackle the problem of domain adaptation, each fold consists of exclusive recording location identifiers. $K$ models are trained using sub-sampled training data, and the entire procedure results in not only the accuracy estimate, but an ensemble of models, as well. The models may then be used to predict labels and confidences for the test data, and fused together either with majority vote or averaging the confidences.

As an example, the convolutional architecture of one of the winning teams consists of four convolutional layers, followed by a layer of Gated Recurrent Units (GRUs) and a softmax layer at the output. 
Without any augmentation, this structure results in a score of 74.66\% on the public leaderboard. If the model is trained separately five times for 5-fold cross-validation, a straightforward majority vote among the 5 different instances yields a significant improvement (78.93\% on public leaderboard) over the baseline approach. 
With mixup augmentation, the accuracy further increases to 79.73\% thus confirming the positive impact of data augmentation techniques in this context, as well.

\subsection{Network Architectures}

\label{ssec:vgg16}

Apart from custom network designs, many teams used networks familiar from image recognition context. Most importantly, the VGG16 network structure \cite{simonyan2014very} is successful with spectral data. Here, the teams use an instance of VGG16  pre-trained with the ImageNet dataset \cite{Imagenet}, but substitute all pre-trained dense layers with randomly initialized fully-connected layers (of dimension 4096, for example). Discarding the original dense layers is preferred, as the convolutional pipeline is not sensitive to the input dimensions (unlike the subsequent dense layers). The data only needs to be replicated along the "color" axis to make the spectrogram dimensions similar to those of a color image; \textit{e.g.,} the spectrogram of size $40\times 501$ is replicated three times to a tensor with dimensions $40\times 501\times 3$.
This architecture alone achieves an accuracy of 78.66\% on the public leaderboard. 




\subsection{Semi-supervised Learning}
\label{ssec:sslr}

Semi-supervised learning is commonly used to integrate large masses of inexpensive unlabeled data with smaller amount of expensive labeled data. In competitions, the natural approach for improving the accuracy is to train a semi-supervised model with the unlabeled test dataset. Although this approach is not applicable in real time systems, it gives valuable insight on how much domain adaptation could improve the accuracy.

A straightforward approach to semi-supervised learning consists of three stages:
\begin{packed_enumerate}
\item Train a model with the training dataset,
\item Predict labels for the test dataset,
\item Add the test samples with the predicted labels into the training set and retrain.
\end{packed_enumerate}
Obviously, one may iterate the above procedure more than once, or add only those samples that were classified with high confidence (\textit{e.g.,} with $>$ 50 \% confidence).
As an example, one of the top teams applied this approach repeatedly three times reaching about 4.5\% improvement compared to a baseline model. 

\begin{table}[t]
\caption{Advanced feature engineering approaches added on top of the baseline model with augmentation (baseline = best result of Table \ref{table:1}).}
\vspace*{0.2cm}
\centering
\begin{tabular}{l||c c}
\hline\hline
\textbf{Method} & \textbf{Public LB} & \textbf{\begin{tabular}[c]{@{}l@{}}Delta from\\  base model\end{tabular}} \\ \hline\hline
Base-line model only & 77.07 \% & 0.0 \\ 
+ Temporal averaging  & 77.60 \% & +0.53 \\ 
+ Background subtraction  &  81.60 \% & +4.53 \\ 
Fusion of the above  &  \textbf{84.13 \%} & +7.07 \\ 
\hline\hline
\end{tabular}%
\label{table:3}
\end{table}

\subsection{Preprocessing}
\label{ssec:twm}

Finally, several manually engineered feature engineering pipelines were discovered. Among the most successful ones are the following.

{\em Temporal averaging}: This technique adds a layer at the beginning of the neural network pipeline to collapse the temporal dimension away: In other words, we average each frequency bin over all time points, transforming the $40\times 501$-dimensional input into $40$-dimensional vector.  By doing so, we emphasize on the influences of frequency bins and completely ignore the temporal changes of sounds. 

\textit{Background subtraction:} This preprocessing step normalizes each frequency bin by subtracting the mean of each sample. As a result, the mean over each 10 second sample becomes zero, emphasizing the temporal patterns of low-energy frequency bins otherwise masked by high-energy bins. 

The effect of these preprocessing steps is studied in Table~\ref{table:3}. The top row shows the accuracy of the baseline model with augmentation and CLR (see Table \ref{table:1}). If we preprocess the data using temporal averaging, we can see minor improvement (+0.53 \%). On the other hand, using background subtraction instead, results in a significant improvement (4.53 \%) over the baseline. Moreover, fusing the three models together (plain baseline, baseline with temporal averaging and baseline with background subtraction) improves the accuracy even further, with a net accuracy increase of 7.07 \%. 


\section{Educational Value}
\label{sec:Education}

The competition was a mandatory part of a graduate level pattern recognition course. Therefore, a valid question is whether the competition adds value to the course and whether the students feel they actually learned something during the course of the competition. 

The overall satisfaction for the course was high. The students are required to submit a numerical evaluation  for general satisfaction of the course organization (grading: 1 = lowest, 5 = highest). The average grade of 110 respondents was 4.35, and none of the students gave a grade less than three. These both indicate that the course in general was very well liked. Of course, the competition is only one part of the course, so we also wanted to investigate the effect of the competition to the overall satisfaction. 

To this aim, we analyzed the verbal feedback given by the students. The verbal feedback is fed into two sections: "What worked well during the course?" and "How would you improve the course"; \textit{i.e.,} the verbal answers are grouped into positive and negative responses. The competition was mentioned in 30.4 \% of the positive responses and in 16.7 \% of the negative ones. Moreover, the negative feedback regarding the competition was exclusively organizational; \textit{e.g.,} regarding team formation, timetables, or GPU availability. 

Qualitatively, the feedback can be categorized into a few broad groups as follows (with some examples translated to English by the authors).

\emph{Scalability.} Unlike traditional project works, the competition is less limited and allows experimentation with an unlimited number of ways. Examples of feedback from this category include the following:
\begin{quote}
\emph{''...the Kaggle competition let those with time and motivation to push themselves and learn as much as possible.''}
\end{quote}
\begin{quote}
\emph{''The competition was an excellent motivator, which does not necessarily require huge amount of time.''}
\end{quote}

\emph{Reflectivity.} An important aspect of student motivation comes from reflecting the skills of an individual against those of the others. In an online competition, a participant receives the feedback immediately and gets an immediate (although noisy) feedback for the performance. Examples of feedback from this category include the following:
\begin{quote}
\emph{''Really, there was a desire to improve the score due to a live comparison with the others.''}
\end{quote}

\emph{Motivation.} Many students felt that the competition is motivating and allows them to learn real hands-on skills, including the following.
\begin{quote}
\emph{''The competition organized during the course was very motivating and at least I learned the most there.''}
\end{quote}

\section{Conclusion}
\label{sec:conclusion}

In this paper, we have described an implementation of a machine learning competition in acoustic scene classification. Primarily, the competition was opened internally for students, but was also open to everyone and did successfully attract participants from all over the world.  As a result, we described a collection of components that contributed to a successful result. 

It is noteworthy, that none of the methods are limited to audio only. Indeed, many approaches (cutout, mixup, etc.) were originally proposed for image data, and some---to the best of our knowledge---not applied to audio before.
In total, the competition attracted 69 teams with altogether over 700 submissions. Roughly one third of the teams were outside the student group, which increases the motivation among students: the competition is not just about one course. Student feedback also suggests that gamification can be a significant part of modern higher level education, while simultaneously progressing  also 
the state of the art in research.

\bibliographystyle{IEEEbib}
{\small\bibliography{refs.bib}}

\end{document}